\documentclass{siamltex}

\usepackage{amssymb,amsmath}
\usepackage{graphicx}
\usepackage{algorithmic}
\usepackage{algorithm}
\usepackage{hyperref}
\usepackage{bm}
\usepackage[bottom]{footmisc}

\newenvironment{packed_enum}{
	\begin{enumerate}
		\setlength{\itemsep}{1pt}
	    \setlength{\parskip}{0pt}
		\setlength{\parsep}{0pt}
}{\end{enumerate}}

\newenvironment{packed_itemize}{
	\begin{itemize}
		\setlength{\itemsep}{1pt}
	    \setlength{\parskip}{0pt}
		\setlength{\parsep}{0pt}
}{\end{itemize}}

\newtheorem{example}[theorem]{Example}
\newtheorem{observation}[theorem]{Observation}
\newtheorem{problem}[theorem]{Problem}

\newcommand{\heading}[1]{\smallbreak\par\noindent\hbox{\bf #1}}

\def\eps{\varepsilon}

\def\O{\mathcal{O}{}}

\def\wlt{\vartriangleleft}

\def\clql{\mathord\curvearrowleft}
\def\clqr{\mathord\curvearrowright}
\def\cp{\textrm{\rm cp}}
\def\LH{\textrm{\rm LH}}
\def\UH{\textrm{\rm UH}}

\def\calI{{\cal I}}

\def\calR{{\cal R}}
\def\calC{{\cal C}}

\def\recog{\textsc{Recog}}
\def\ext{\textsc{RepExt}}
\def\reorder{\textsc{Reorder}}

\def\simrep{\textsc{SimRep}}

\def\int{\hbox{\rm \sffamily INT}}
\def\pint{\hbox{\rm \sffamily PROPER\ INT}}
\def\uint{\hbox{\rm \sffamily UNIT\ INT}}

\def\path{\hbox{\rm \sffamily P-in-T}}
\def\chor{\hbox{\rm \sffamily CHOR}}
\def\ca{\hbox{\rm \sffamily CIRCULAR-ARC}}

\def\cNP{\hbox{\rm \sffamily NP}}

\def\cFPT{\hbox{\rm \sffamily FPT}}

\def\computationproblem#1#2#3{
	\medskip
	\begin{center}
	\begin{tabular}{rp{8.5cm}}
	{\sc Problem:\enspace}&#1\\
	{\sc Input:\enspace}&#2\\
	{\sc Question:\enspace}&#3\\
	\end{tabular}
	\end{center}
	\medskip
}

\title{Extending Partial Representations\\of Interval Graphs\thanks{%
A conference version of this paper appeared in TAMC 2011~\cite{kkv}.  Supported by ESF Eurogiga
project GraDR as GA\v{C}R GIG/11/E023. The first two authors are also supported by Charles University as GAUK 196213.}}
\author{P.~Klav\'{I}k\footnotemark[2]
		\and J.~Kratochv\'Il\footnotemark[3]
		\and Y.~Otachi\footnotemark[4]
		\and T.~Saitoh\footnotemark[5]
		\and T.~Vysko\v{c}il\footnotemark[3]}

\begin{document}
\maketitle

\renewcommand{\thefootnote}{\fnsymbol{footnote}}
\footnotetext[2]{Computer Science Institute, Faculty of Mathematics and Physics, Charles University,
Malostransk{\'e} n{\'a}m{\v e}st{\'\i} 25, 118 00 Prague, Czech Republic. E-mail:
\texttt{klavik@iuuk.mff.cuni.cz}.}
\footnotetext[3]{Department of Applied Mathematics, Faculty of Mathematics and Physics, Charles
University, Malostransk{\'e} n{\'a}m{\v e}st{\'\i} 25, 118 00 Prague, Czech Republic. E-mails:
\texttt{honza@kam.mff.cuni.cz} and \texttt{whisky@kam.mff.cuni.cz}.}
\footnotetext[4]{School of Information Science, Japan Advanced Institute of Science and Technology.
Asahidai 1-1, Nomi, Ishikawa 923-1292, Japan. Email: \texttt{otachi@jaist.ac.jp}}
\footnotetext[5]{Graduate School of Engineering, Kobe University, Rokkodai 1-1, Nada, Kobe,
657-8501, Japan.  E-mail: \texttt{saitoh@eedept.kobe-u.ac.jp}}
\renewcommand{\thefootnote}{\arabic{footnote}}

\begin{abstract}
Interval graphs are intersection graphs of closed intervals of the real-line. The well-known
computational problem, called \emph{recognition}, asks whether an input graph $G$ can be
represented by closed intervals, i.e., whether $G$ is an interval graph. There are several
linear-time algorithms known for recognizing interval graphs, the oldest one is by Booth and Lueker
[J.~Comput. System Sci., 13 (1976)] based on PQ-trees.

In this paper, we study a generalization of recognition, called \emph{partial representation
extension}. The input of this problem consists of a graph $G$ with a partial representation $\calR'$
fixing the positions of some intervals. The problem asks whether it is possible to place the
remaining interval and create an interval representation $\calR$ of the entire graph $G$ extending
$\calR'$.  We generalize the characterization of interval graphs by Fulkerson and Gross
[Pac.~J.~Math., 15 (1965)] to extendible partial representations. Using it, we give a linear-time
algorithm for partial representation extension based on a reordering problem of PQ-trees.
\end{abstract}

\section{Introduction}

One of the fundamental themes of mathematics is studying relations between mathematical objects and
their representations. For graph theory, the study of graph representations and graph drawing is as
old as the study of graphs themselves. A widely studied type of graph representations are
intersection representation which encode edges by intersections of sets. An \emph{intersection
representation} $\calR$ of a graph $G$ assigns a collection of sets $\bigl\{R_v \mid v \in
V(G)\bigr\}$ such that $uv \in E(G)$ if and only if $R_u \cap R_v \ne \emptyset$. Since every graph
has an intersection representation~\cite{every_graph_is_an_intersection_graph}, interesting graph
classes are obtained by restricting the representing sets to some nice class of, say, geometrical
objects, e.g., continuous curves in plane, chords of a circle, convex sets, etc. For overview of
these classes, see books~\cite{agt,tig,egr}.

The most famous are \emph{interval graphs} (\int) which are intersection graphs of closed intervals
of the real line. It is one of the oldest classes of graphs, introduced by Haj\'os~\cite{hajos}
already in 1957.  Interval graphs have many useful theoretical properties, for example they are
perfect and related to path decompositions. In many cases, very hard combinatorial problems are
polynomially solvable for interval graphs~\cite{recog_chordal_graphs}; e.g., maximum clique,
$k$-coloring, maximum independent set, etc. Also, interval graphs naturally appear in many
applications concerning biology, psychology, time scheduling, and archaeology; see for
example~\cite{roberts_discrete_models, stoffers,benzer_interval_graphs}.

\heading{Partial Representation Extension.}
For a fixed class $\calC$, there is a natural well-studied problem called \emph{recognition}. Given
a graph $G$, we ask whether $G$ belongs to $\calC$. We denote this problem by $\recog(\calC)$. For
interval graphs, there are several algorithms solving $\recog(\int)$ in linear
time~\cite{PQ_trees,LBFS_int,finding_lb_graphs}. Further, interval graphs have nice mathematical
characterizations~\cite{maximal_cliques,lb_graphs} which are foundations of these algorithms. 

In this paper, we introduce a natural generalization of recognition called \emph{partial
representation extension}. A \emph{partial representation} $\calR'$ of $G$ is a representation of an
induced subgraph $G'$ of $G$. The vertices of $G'$ are called \emph{pre-drawn}. A representation
$\calR$ of $G$ \emph{extends} $\calR'$ if it assigns the same sets to the vertices of $G'$, i.e.,
$R_v = R'_v$ for every $v \in V(G')$. Partial representation extension is the following decision
problem:

\computationproblem
{Partial representation extension -- $\ext(\calC)$}
{A graph $G$ and a partial representation $\calR'$.}
{Is there a representation $\calR$ of $G$ extending $\calR'$?}

\begin{figure}[t!]
\centering
\includegraphics{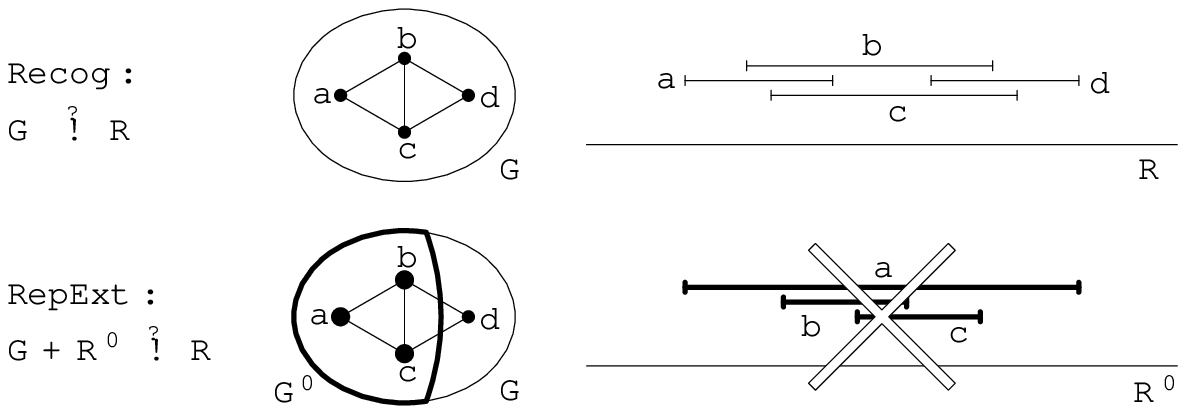}
\caption{The graph $G$ is an interval graph, but the partial representation $\calR'$ is not
extendible.}
\label{fig:comparison}
\end{figure}

\noindent Figure~\ref{fig:comparison} illustrates the difference between the recognition problem and
the partial representation extension problem.

In this paper, we initiate the study of partial representation extension with interval graph.  We
have two reasons to choose interval graphs. First, this class is one of the oldest and most
understood. As an evidence of its popularity, Web of Knowledge lists more than 300 papers with
the words ``interval graphs'' in the title. Second, there are many structural results and techniques
known for interval graphs. Namely we can use PQ-trees~\cite{PQ_trees} which combinatorially describe all
interval representations of an interval graph. This way we discover the most important properties
of partial representation extension, applicable to other more complex graph classes, without dealing
with technical details. In particular, we show that a good understanding of the structure of all
representations is essential to solve the problem; unlike recognition for which any representation
has to be found.

We give the following main algorithmic result, where sorted representations are defined in the end
of this section:

\medskip

\begin{theorem} \label{thm:ext_int}
If the partial representation is given sorted from left to right, then the problem $\ext(\int)$ can be
solved in time $\O(n+m)$, where $n$ is the number of vertices and $m$ is the number of edges.
\end{theorem}

\medskip

Fulkerson and Gross~\cite{maximal_cliques} proved in 1965 the following characterization. A~graph is
an interval graphs if and only if there exists an ordering of its maximal cliques such that for each
vertex the cliques containing this vertex appear consecutively in this ordering.  Intervals of the
real line have the Helly property, so all intervals representing one maximal clique have a common
intersection. In this intersection, we choose one point which we call a clique-point.  The
considered ordering is the left-to-right ordering of the chosen clique-points.

\begin{figure}[t!]
\centering
\includegraphics{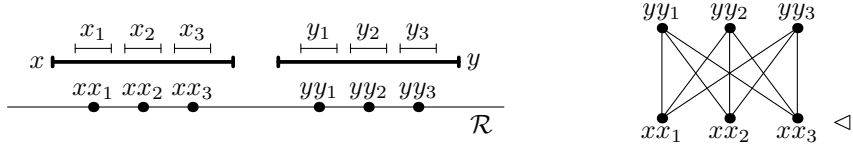}
\caption{An interval graph consisting of two stars with pre-drawn central vertices. One of the
extending representations is depicted on the left.  Any extending representation places all maximal
cliques containing $x$ on the left of the maximal cliques containing $y$. Therefore the ordering of
the maximal cliques has to extend the partial ordering $\wlt$, depicted by the Hasse diagram on the
right.}
\label{fig:quadratic_example}
\end{figure}

We generalize this result and characterize extendible partial representations.  The partial
representation gives a partial ordering $\wlt$ which has to be extended by the ordering of the
maximal cliques of any extending representation; see Figure~\ref{fig:quadratic_example} for an
example. Our characterization of extendible instances says that the constraints posed by $\wlt$ are
not only necessary, but also sufficient. 

The algorithm of Theorem~\ref{thm:ext_int} tests this characterization. The data structure called
PQ-tree combinatorially describes all orderings of the maximal cliques yielding interval
representations. We test whether this tree can be reordered according to $\wlt$. By applying several
tricks, we can test this for a specific type of partial orderings called interval orders in linear
time.

\heading{Previous results of \ext.}
Concerning previous results, the conference version of this paper~\cite{kkv} shows that
interval representations can be extended in time $\O(n^2)$ and proper interval representations can
be extended in time $\O(nm)$. For interval graphs, Bl\"asius and Rutter~\cite{blas_rutter} improve
this result to time $\O(n+m)$. They reduce it to a more general problem called \emph{simultaneous
representations} which they solve using simultaneous PQ-trees. This framework also applies to other
problems such as simultaneous planar embeddings. Consequently their algorithm is quite involved and
gives no understanding of partial representation extension.

Compared to the algorithm of Bl\"asius and Rutter~\cite{blas_rutter}, there are three main points
why our linear-time algorithm for partial representation extension is interesting.
\begin{packed_enum}
\item Our algorithm is much simpler and easier to implement.
\item Some of our understanding and techniques can be applied to more complicated classes, for which
either the simultaneous representations problem is still open, or the relation with partial
representation extension is lost. For instance, the simultaneous representation problem is
polynomially solvable for chordal graphs~\cite{jampani2}, but partial representation extension is
\cNP-complete~\cite{kkos}.
\item Our structural understanding can be used to obtain further results. Namely, Balko et
al.~\cite{bko} proved that our algorithm can be simply modified to a more general problem called
\emph{bounded representation}. To the best of our knowledge, there are no relations between bounded
and simultaneous representations. Further, the paper~\cite{ks} uses our characterization of
extendible instances to describe minimal obstructions for extendibility of partial interval
representations. This generalizes the result of Lekherkerker and Boland~\cite{lb_graphs}
which describes all minimal graphs which are not interval graphs.
\end{packed_enum}

Every interval graph has a representation in which every endpoint is placed at an integer position.
We note that extending such representations, again by integer endpoints, is an \cNP-complete
problem~\cite{kkos}.

The paper~\cite{kkorssv} improves~\cite{kkv} by giving a linear-time algorithm for proper interval
graphs, and it also solves an open problem of~\cite{kkv} by giving an almost quadratic-time
algorithm for unit interval graphs. These two results might seem surprising in the
context of Robert's Theorem~\cite{roberts_theorem} which states that these two classes are
equal, i.e, $\pint = \uint$. But the partial representation extension problems distinguish them.
In the case of proper interval representations, a partial representation just prescribes some partial
ordering of the endpoints of the intervals. But a partial unit interval representation gives in
addition precise rational positions. The algorithm of~\cite{kkorssv} for unit interval graphs is
based on linear programming and new structural results.

The paper~\cite{kkkw} gives polynomial-time algorithms for permutation and function graphs.
The paper~\cite{kkos} studies several possible versions of the problem for chordal graphs (in the
setting of intersection graphs of subtrees of a tree) and shows that almost all of them are
\cNP-complete. Concerning circle graphs, a polynomial-time algorithm is given by Chaplick et
al.~\cite{cfk}. It is based on new structural results which describe via split decomposition all
possible representations of circle graphs.

For planar graphs, Angelini et al.~\cite{angelini} show that partial planar embeddings can be
extended in linear time. Well-known F\'ary's Theorem states that every planar graph has a
straight-line embedding. But it is \cNP-complete to decide whether a partial straight-line embedding
can be extended to a straight-line embedding of the entire graph~\cite{patrignani}.

\heading{Motivation for \ext.} To solve the recognition problem, an arbitrary representation can be
constructed. Solving partial representation extension is harder, and better understanding of the
structure of all possible representations seems to be necessary. This is a desirable property since
one is forced to improve the structural understanding of the studied classes to solve this problem;
and this structural understanding can be later applied in attacking other problems.

The structure of all representations of interval graphs is already well understood~\cite{PQ_trees},
so for our algorithm we just use this structure. On the other hand, the papers~\cite{kkorssv,cfk}
build completely new structural results for unit interval and circle graphs which might be of
independent interest.

Partial representation extension belongs to a larger group of restricted representation problems. In
these problems, one asks whether there exists a representation satisfying some additional
constraints. We define the simultaneous representations problem in
\S\ref{sec:simultaneous_representations}. The bounded representation problem gives two restricting
intervals $A_v$ and $B_v$ to each vertex $v \in V(G)$.  The task is to find a representation which
places one endpoint of the interval $R_v$ into $A_v$ and the other one to $B_v$. So it is a
relaxation of the partial representation extension since we can move endpoints of $R_v$ slightly. There are
several similar problems studied for intervals graphs; for example
see~\cite{int_prescribed_lengths,int_lengths_intersections}.

\heading{Sorted Representations.}
To obtain the linear-time algorithm, we need some reasonable assumption on a partial representation
which is given by the input. Similarly, most of the graph algorithms cannot achieve better running
time than $\O(n^2)$ if the input graph is given by an adjacency matrix instead of a list of
neighbors for each vertex.

We say that a partial representation is \emph{sorted} if it gives all (left and right) endpoints of
the pre-drawn intervals sorted from left to right. We assume that the input partial representation
is given sorted. If this assumption is not satisfied, the algorithm needs additional time $\O(k \log
k)$ to sort the partial representation where $k$ is the number of pre-drawn intervals. We note that
Bl\"asius and Rutter~\cite{blas_rutter} need the same assumption for their linear-time algorithm.

\heading{Structure.}
This paper is structured as follows.

In \S\ref{sec:pq_tree_reordering}, we define a PQ-tree which is a data structure describing all
representations of an interval graph. We introduce a reordering problem asking whether the leaves
of a PQ-tree can be reordered according to some partial ordering $\wlt$. We give two algorithm for
this problem: one for general partial orderings, and another faster one for interval orders which
are partial orderings represented by collections of open intervals. 

In \S\ref{sec:extension_algorithm}, we build a bridge between PQ-trees and interval graphs. We
derive characterization of extendible instances saying a partial representation is extendible if and
only if the PQ-tree can be reordered according to some interval ordering $\wlt$. The main algorithm
of Theorem~\ref{thm:ext_int} just computes this interval ordering and applies the second reordering
algorithm as a subroutine.

In \S\ref{sec:simultaneous_representations}, we discuss connections with the simultaneous
representations problems and show that Theorem~\ref{thm:ext_int} gives an \cFPT\ algorithm for the
simultaneous representations problem of interval graphs. We conclude in \S\ref{sec:open_problem}
with the current major open problem of partial representation extension.

\section{PQ-trees and the Reordering Problem} \label{sec:pq_tree_reordering}

To describe PQ-trees, we start with a motivational problem. An input of the \emph{consecutive
ordering problem} consists of a set $E$ of elements and restricting sets $S_1, S_2, \dots, S_k$. The
task is to find a (linear) ordering of $E$ such that every $S_i$ appears consecutively (as one
block) in this ordering.

\begin{example} \label{consecutive_ordering_example}
Consider the elements $E = \{a,b,c,d,e,f,g,h\}$ and the restricting sets $S_1 = \{a,b,c\}$, $S_2 =
\{d,e\}$, and $S_3 = \{e,f,g\}$. For instance, the orderings $\textit{abcdefgh}$ and
$\textit{fgedhacb}$ are feasible. On the other hand, the orderings
$\textit{\underline{ac}defg\underline{b}h}$ (violates $S_1$) and
$\textit{d\underline{ef}h\underline{g}abc}$ (violates $S_3$) are not feasible.
\end{example}

\heading{PQ-trees.} A PQ-tree is a tree structure invented by Booth and Lueker~\cite{PQ_trees} for
solving the consecutive ordering problem efficiently. Moreover, it stores all feasible orderings for
a given input.

\begin{figure}[b!]
\centering
\includegraphics{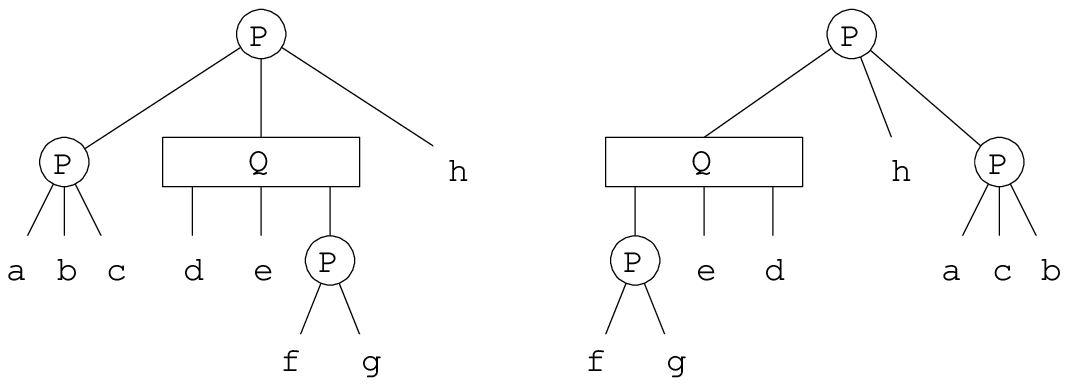}
\caption{PQ-trees representing orderings $\textit{abcdefgh}$ and $\textit{fgedhacb}$.}
\label{PQ_trees}
\end{figure}

The leaves of the tree correspond one-to-one to the elements of $E$. The inner nodes are of two
types: The \emph{P-nodes} and the \emph{Q-nodes}. The tree is rooted and an order of the children of
every inner node is fixed. Also we assume that each inner node has at least two children. A PQ-tree
$T$ represents one ordering $<_T$, given by the ordering of the leaves from left to right, see
Figure~\ref{PQ_trees}.

To obtain other feasible orderings, we can reorder children of inner nodes. The children of a P-node
can be reordered in an arbitrary way.  On the other hand, we can only reverse the order of the
children of a Q-node.  We say that a tree $T'$ is a \emph{reordering} of $T$ if it can be created
from $T$ by applying several reordering operations. Two trees are \emph{equivalent} if one is
a reordering of the other. For example, the trees in Figure~\ref{PQ_trees} are equivalent. Every
equivalence class of PQ-trees corresponds to all the orderings feasible for a fixed family of
equivalent input sets. The equivalence class of the PQ-trees in Figure~\ref{PQ_trees} corresponds to
the input sets in Example~\ref{consecutive_ordering_example}.

For the purpose of this paper, we only need to know that a PQ-tree can be constructed in time
$\O(e+k+t)$ where $e$ is the number of elements of $E$, $k$ is the number of restricting sets and
$t$ is the sum of cardinalities of restricting sets. Booth and Lueker~\cite{PQ_trees} prove their
existence and describe details of their construction.

\subsection{The Reordering Problem for General Orderings} \label{subsec:general_order}

Suppose that $T$ is a PQ-tree and we have a partial ordering $\wlt$ of its elements (leaves). We say
that a reordering $T'$ of the PQ-tree $T$ is \emph{compatible} with $\wlt$ if the ordering $<_{T'}$
extends $\wlt$, i.e., $a \wlt b$ implies $a <_{T'} b$.

\computationproblem
{The reordering problem -- $\reorder(T,\wlt)$}
{A PQ-tree $T$ and a partial ordering $\wlt$.}
{Is there a reordering $T'$ of $T$ compatible with $\wlt$?}

\heading{Local Solutions.} A PQ-tree defines some hierarchical structure on its elements. A
\emph{subtree} of a PQ-tree consists of one inner node and all its successors.

\begin{observation} \label{obs:consecutive_subtrees}
Let $S$ be a subtree of a PQ-tree $T$. Then the elements of $E$ contained in $S$ appear
consecutively in $<_T$. 
\end{observation}

We start with a lemma which states the following: If we can solve the problem locally (inside of
some subtree), then this local solution is always correct; either there exists no solution of the
problem at all, or our local solution can be extended to a solution for the whole tree.

\begin{lemma} \label{lem:local_reordering}
Let $S$ be a subtree of a PQ-tree $T$. If $T$ can be reordered compatibly with $\wlt$, then every local
reordering of the subtree $S$ compatible with $\wlt$ can be extended to a reordering of the whole tree
$T$ compatible with $\wlt$.
\end{lemma}

\begin{proof}
Let $T'$ be a reordering of the whole PQ-tree $T$ compatible with $\wlt$. According to
Observation~\ref{obs:consecutive_subtrees}, all elements contained in $S$ appear consecutively in
$<_{T'}$.  Therefore, we can replace this local ordering of $S$ by any other local ordering of $S$
satisfying all constraints given by $\wlt$. We obtain another reordering of the whole tree $T$ which
is compatible with $\wlt$ and extends the prescribed local ordering of $S$.
\end{proof}

\heading{The Algorithm.} We describe the following algorithm for $\reorder(T,\wlt)$:

\begin{proposition} \label{prop:general_reorder}
The problem $\reorder(T,\wlt)$ can be solved in time $\O(e+m)$, where $e$ is the number of elements
and $m$ is the number of comparable pairs in $\wlt$.
\end{proposition}

\begin{proof}
The algorithm is based on the following greedy procedure. We represent the ordering $\wlt$ by a
digraph having $m$ edges.  We reorder the nodes from the bottom to the root and modify the digraph by
contractions. When we finish reordering a subtree, the order is fixed and never changed in the
future; by Lemma~\ref{lem:local_reordering}, either this local reordering will be extendible, or
there is no correct reordering of the whole tree at all. When we finish reordering a subtree, we
contract the corresponding vertices in the digraph. We process a node of the PQ-tree when all its
subtrees are already processed and their digraphs are contracted to single vertices.

\begin{figure}[t!]
\centering
\includegraphics[scale=0.95]{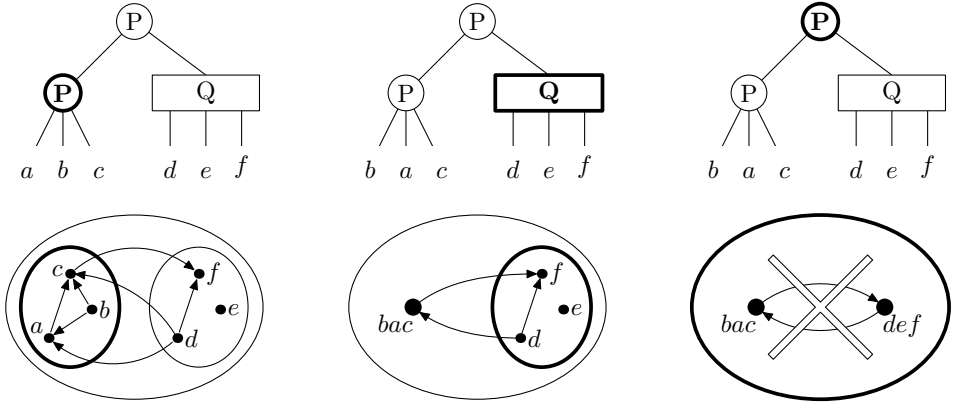}
\caption{We show from left to right an example how the reordering algorithm works. First, we reorder
the highlighted P-node on the left. The subdigraph induced by $a$, $b$ and $c$ has the topological
sort $b \rightarrow a \rightarrow c$. We contract these vertices into the vertex $bac$. Next, we
keep the order of the highlighted Q-node and contract its children into the vertex $def$. When we
reorder the root P-node, the algorithm finds a cycle between $bac$ and $def$, and outputs ``no''.
Notice that the original digraph $\wlt$ is acyclic, just not compatibly with the structure of the
PQ-tree.}
\label{pq_tree_reordering}
\end{figure}

For a P-node, we check whether the subdigraph induced by the vertices corresponding to the children
of the P-node is acyclic. If it is acyclic, we reorder the children according to any topological
sort of the subdigraph. Otherwise, there exists a cycle, no feasible ordering exists and the algorithm returns
``no''. For a Q-node, there are two possible orderings. We just need to check whether one of them is
feasible. For an example, see Figure~\ref{pq_tree_reordering}.

We need to argue the correctness. The algorithm processes the tree from the bottom to the top. For every
subtree $S$, it finds some reordering of $S$ compatible with $\wlt$.  If no such reordering of $S$
exists, the whole tree $T$ cannot be reordered according to $\wlt$. If a reordering of $S$
exists, it is correct according to Lemma~\ref{lem:local_reordering}.

The algorithm can be implemented in linear time with respect to the size of the PQ-tree and the~partial
ordering $\wlt$ which is $\O(e+m)$. Each edge of the digraph $\wlt$ is processed exactly once before
it is contracted.
\end{proof}

\begin{algorithm}[b!]
\caption{Reordering a PQ-tree -- $\reorder(T,\wlt)$} \label{alg:general_reorder}
\begin{algorithmic}[1]
\REQUIRE A PQ-tree $T$ and a partial ordering $\wlt$.
\ENSURE A reordering $T'$ of $T$ such that $<_{T'}$ extends $\wlt$ if it exists.
\medskip

\STATE Construct the digraph of $\wlt$.
\medskip
\STATE Process the nodes of $T$ from the bottom to the root:
\FOR{a processed node $N$}
	\STATE Consider the subdigraph induced by the children of $N$.
	\IF{the node $N$ is a P-node}
		\STATE Find a topological sort of the subdigraph.
		\STATE If it exists, reorder $N$ according to it, otherwise output ``no''.
	\ELSIF{the node $N$ is a Q-node}
		\STATE Test whether the current ordering or its reversal are compatible\\with the subdigraph.
		\STATE If at least one is compatible, reorder the node, otherwise output ``no''.
	\ENDIF
	\STATE Contract the subdigraph into a single vertex.
\ENDFOR
\medskip
\RETURN A reordering $T'$ of $T$.
\end{algorithmic}
\end{algorithm}

We note that the described algorithm works even for a general relation $\wlt$. For example, $\wlt$
does not have to be transitive (as in the example in Figure~\ref{pq_tree_reordering}) or even
acyclic (but in such a case, of course, no solution exists). A pseudocode is given in
Algorithm~\ref{alg:general_reorder}.

\subsection{The Reordering Problem for Interval Orders} \label{subsec:interval_order}

In this section, we establish a faster algorithm for the reordering problem for a special type of
partial orderings called interval orders. We first define them.

\begin{figure}[t!]
\centering
\includegraphics{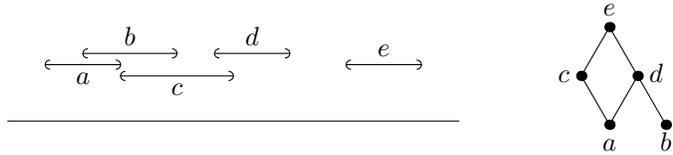}
\caption{On the left, a collection of open intervals. On the right, the Hasse diagram of the
interval order $\wlt$ represented by these intervals.}
\label{fig:example_of_interval_order}
\end{figure}

Let $E$ be a set and let $\{I_a = (\ell_a,r_a) \mid a \in E\}$ be a collection of open
intervals.\footnote{For the purpose of \S\ref{sec:extension_algorithm}, we allow empty intervals
with $\ell_v = r_v$.} Then these intervals \emph{represent} the following partial ordering $\wlt$ on
$E$. If two intervals $I_a$ and $I_b$ do not intersect, then one is on the left and the other is on
the right. Therefore we naturally ordered intervals as they appear from left to right. Formally, for
$a,b \in E$, we put $a \wlt b$ if and only if $r_a \le \ell_b$. A partial ordering of $E$ is called
an \emph{interval order} if there exists a collection of intervals representing this ordering in
this way.  See Figure~\ref{fig:example_of_interval_order} for an example.

Both interval graphs and interval orders are represented by collections of intervals, and indeed
they are closely related~\cite{fishburn}. The study of interval orders has the following motivation.
Suppose that the elements of $E$ correspond to events and each interval describes when one event can
happen in the timeline.  If $a \wlt b$, we know for sure that the event $a$ happened before the
event $b$. If two intervals intersect, we do not have any information about the order of the
corresponding events. Nevertheless, for purpose of this paper, we only need to know the definition
of interval orders. For more information, see the survey~\cite{trotter}.

\heading{Faster Reordering of PQ-trees.}
Let $e$ be the number of elements of $E$ and let $\wlt$ be an interval order of $E$ represented by
$\{I_a \mid a \in E\}$. We assume the representation is sorted which means that we know the order of
all endpoints of the intervals from left to right. We show that for such $\wlt$ we can solve
$\reorder(T,\wlt)$ faster:

\begin{proposition} \label{prop:interval_reorder}
If $\wlt$ is an interval order given by a sorted representation, we can solve the problem
$\reorder(T,\wlt)$ in time $\O(e)$ where $e$ is the number of elements of~$T$.
\end{proposition}

For the following, let $\lessdot$ be the linear ordering of the endpoints $\ell_e$ and $r_e$ of the
intervals according to their appearance from left to right in the representation.  To ensure that $a
\wlt b$ if and only if $r_a \lessdot \ell_b$, we need to deal with endpoints sharing position. For
them, we place in $\lessdot$ first the right endpoints (ordered arbitrarily) and then the left
endpoints (again ordered arbitrarily). For a sorted representation, this ordering $\lessdot$ can be
computed in time $\O(e)$. For example in Figure~\ref{fig:example_of_interval_order} we get
$$\ell_a \lessdot \ell_b \lessdot r_a \lessdot \ell_c \lessdot r_b \lessdot \ell_d
\lessdot r_c \lessdot r_d \lessdot \ell_e \lessdot r_e.$$

The general outline of the algorithm is exactly the same as before. We process the nodes of the
PQ-tree from the bottom to the root and reorder them according to the local constraints. Using the interval
representation of $\wlt$, we can implement all steps faster than before.

Informally speaking, the main trick is that we do not construct the digraph explicitly. Instead, we
just work with sets of intervals corresponding to subtrees and compare them with respect to $\wlt$
fast. When we process a node, its children correspond to sets $\calI_1,\dots,\calI_k \subseteq E$
we already processed before. We test efficiently in time $\O(k)$ whether we can reorder these $k$
subtrees according to $\wlt$. If it is not possible, the algorithm stops and outputs ``no''. If the
reordering succeeds, we put all the sets together $\calI = \calI_1 \cup \calI_2 \cup \cdots \cup
\calI_k$, and proceed further. We now describe everything in details.

\heading{Comparing Subtrees.} Let $\calI_1$ and $\calI_2$ be sets of intervals. We say $\calI_1 \wlt
\calI_2$ if there exist $a \in \calI_1$ and $b \in \calI_2$ such that $a \wlt b$.  We want to show
that using the interval representation and some precomputation, we can test whether $\calI_1 \wlt
\calI_2$ in constant time.  The following lemma states that we just need to compare the
``left-most'' interval of $\calI_1$ with the ``right-most'' interval of $\calI_2$.

\begin{lemma} \label{lem:comparing_lemma}
Suppose that $a \wlt b$ for $a \in \calI_1$ and $b \in \calI_2$. Then for every $a' \in \calI_1$ with
$r_{a'} \lessdot r_a$ and every $b' \in \calI_2$ with $\ell_b \lessdot \ell_{b'}$, it holds that $a'
\wlt b'$.  
\end{lemma}

\begin{figure}[t!]
\centering
\includegraphics{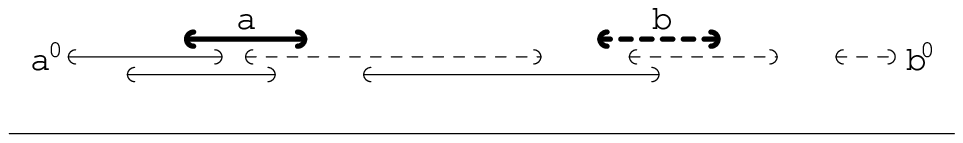}
\caption{The normal intervals belong to $\calI_1$ and the dashed intervals belong to $\calI_2$.
If $a \wlt b$, then also $a' \wlt b'$.}
\label{fig:fast_comparison}
\end{figure}

\begin{proof}
From the definition, $a \wlt b$ if and only if $r_a \le \ell_b$. We have $r_{a'} \lessdot r_a
\lessdot \ell_b \lessdot \ell_{b'}$, and thus $a' \wlt b'$. See Figure~\ref{fig:fast_comparison}.
\end{proof}

Using this lemma, we just need to compare $a$ having the left-most $r_{a}$ to $b$ having the
right-most $\ell_{b}$ since $\calI_1 \wlt \calI_2$ if and only if $a \wlt b$.

\begin{figure}[b!]
\centering
\includegraphics{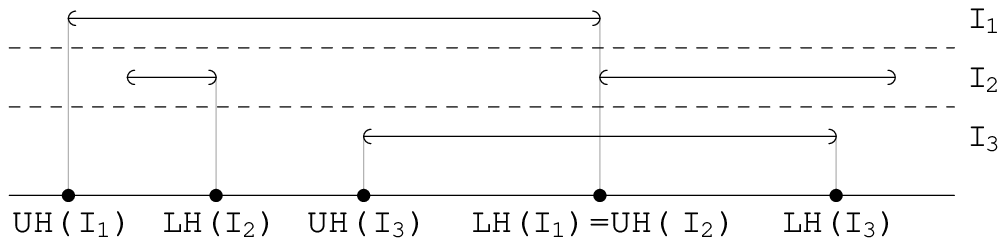}
\caption{The handles for sets $\calI_1$, $\calI_2$ and $\calI_3$. We have
$\UH(\calI_1) \lessdot \LH(\calI_2) \lessdot \UH(\calI_3) \lessdot \LH(\calI_1)
\lessdot \UH(\calI_2) \lessdot \LH(\calI_3)$.
According to~(\ref{eq:fast_comparison}), we get $\calI_1 \wlt \calI_2$, $\calI_2 \wlt \calI_3$, and
$\calI_1 \not\wlt \calI_3$; so the relation $\wlt$ on sets of intervals is not necessarily
transitive.}
\label{fig:order_of_handles}
\end{figure}

To simplify the description, these special endpoints of intervals used for comparisons are called
\emph{handles}. More precisely, for a set of intervals $\calI$, we define a \emph{lower handle}
and an \emph{upper handle}:
\begin{equation}
\LH(\calI) = \min\{ r_x \mid x \in \calI\}
\qquad\textrm{and}\qquad
\UH(\calI) = \max\{ \ell_x \mid x \in \calI\}.
\end{equation}
We note that $\LH(\calI) \lessdot \UH(\calI)$ if $\calI$ is not a clique. Using handles, we can
compare sets of intervals fast. According to Lemma~\ref{lem:comparing_lemma}, we have:
\begin{equation} \label{eq:fast_comparison}
\calI_1 \wlt \calI_2\qquad\text{if and only if}\qquad\LH(\calI_1) \lessdot \UH(\calI_2).
\end{equation}
For an example, see Figure~\ref{fig:order_of_handles}.

So throughout the algorithm, we efficiently compute these handles for each processed subtree, and we
do not need to remember which specific intervals are contained in the subtree. The handles serve in
the same manner as the contraction operation of digraphs in the proof of
Proposition~\ref{prop:general_reorder}.

\heading{Reordering Nodes.} We describe fast reordering of the children of a processed node using
the handles.  Let $\calI_1,\dots,\calI_k$ be the sets of intervals corresponding to the subtrees defined
by the children of this node. Suppose that we know their handles and have them ordered according to
$\lessdot$ as in Figure~\ref{fig:order_of_handles}. Let $\widetilde\lessdot$ be the ordering
$\lessdot$ restricted to the handles of $\calI_1,\dots,\calI_k$. 

A linear ordering $<$ of the sets $\calI_1,\dots,\calI_k$ is called a \emph{topological sort} if
$\calI_i \wlt \calI_j$ implies $\calI_i < \calI_j$ for every $i \ne j$.
An element $\calI_j$ is \emph{minimal} if there is no $\calI_i$ such that $\calI_i \wlt \calI_j$.
We use minimal elements to characterize all topological sorts. For every topological sort $1 <
\cdots < k$, the $\ell$-th element restricted to $\{\ell,\ell+1,\dots,k\}$ is minimal.  We describe
this classical characterization in details since it is important for our algorithm.

Every topological sort can be constructed as follows. We repeatedly detect all minimal element
$\calI_i$ and always pick one of them. (For different choices we get different topological sorts).
We stop when all elements are placed in the topological sort. If in some step no minimal element
exists, we also know that no topological sort exists.

The following lemma describes minimal elements in terms of the ordering $\widetilde\lessdot$:

\begin{lemma} \label{lem:minimal_elements}
Let $\calI_j$ be an element. It is a minimal element if and only if there is no lower handle
$\LH(\calI_i)$ for $i \ne j$ such that $\LH(\calI_i) \widetilde\lessdot \UH(\calI_j)$.
\end{lemma}

\begin{proof}
According to~(\ref{eq:fast_comparison}), $\calI_i \wlt \calI_j$ if and only if $\LH(\calI_i)
\widetilde\lessdot \UH(\calI_j)$. If there is no such $\calI_i$, then $\calI_j$ is minimal.
\end{proof}

We can use this lemma to identify all minimal elements:
\begin{packed_itemize}
\item If the ordering $\widetilde\lessdot$ starts with two lower handles $\LH(\calI_i)$ and
$\LH(\calI_j)$, there exists no minimal element. The reason is that all upper handles are larger,
and so both $\calI_i$ and $\calI_j$ are smaller than everything else; specifically, we get $\calI_i
\wlt \calI_j \wlt \calI_i$.
\item If the first element of the ordering $\widetilde\lessdot$ is $\LH(\calI_i)$ then $\calI_i$ is
the unique candidate for a minimal element.  We just need to check whether there is some other
$\LH(\calI_j)$ smaller than $\UH(\calI_i)$, and if so, no minimal element exists.\footnote{This can be
done in constant time if we remember in each moment the positions of the two left-most lower handles in
the ordering, and update this information after removing one of them from $\widetilde\lessdot$.}
\item If $\widetilde\lessdot$ starts with a consecutive group of upper handles, we have several
candidates for a minimal element. First, all $\calI_i$'s of these upper handles are minimal
elements. Second, if the lower handle following the group of upper handles is $\LH(\calI_j)$, then
$\calI_j$ is a candidate for a minimal element.  As above, $\calI_j$ is minimal if there is no other
lower handle smaller than $\UH(\calI_j)$.
\end{packed_itemize}
When constructing a topological sort, we remove the handles of the picked minimal elements $\calI_i$
from $\widetilde\lessdot$ and append $\calI_i$ to the sort. 

For a P-node, we just need to find any topological sort by repeated removing of minimal elements in
any way.  For a Q-node, we test whether the current ordering or its reversal are topological sorts.
We iterate through each of the two prescribed orderings, check whether each element is a minimal
element, and then remove its handles from $\widetilde\lessdot$. In both cases, if we find a correct
topological sort, we use it reorder the children of the node. Otherwise, the reordering is not
possible and the algorithm outputs ``no''. We are able to do the reordering of the node in time
$\O(k)$.

\heading{The Algorithm.}
We are ready to show that our algorithm allows to find a reordering of the PQ-tree $T$
compatible with an interval order $\wlt$ with a sorted representation in time $\O(e)$:

\begin{proof}[Proposition~\ref{prop:interval_reorder}]
We first deal with details of the implementation. We precompute the handles for every set of
intervals corresponding to a subtree of an inner node of $T$. For each leaf, the handles
are the endpoints. We process the tree from the bottom to the root. Suppose that we have
an inner node corresponding to the set $\calI$ of intervals and it has $k$ children corresponding to
$\calI_1,\dots,\calI_k$ for which we already know their handles. Then we calculate the handles of
$\calI$ using
\begin{equation} \label{eq:handle_computation}
\LH(\calI) = \min \bigl\{\LH(\calI_i)\bigr\}
\qquad\textrm{and}\qquad
\UH(\calI) = \max\bigl\{\UH(\calI_i)\}.
\end{equation}
This can clearly be computed in time $\O(e)$, and we also note for each endpoint a list of nodes for
which it is a handle. Using these list, we can sweep the sorted representation and compute all
orderings $\widetilde\lessdot$ for all inner nodes of $T$, again in $\O(e)$ time.

\begin{algorithm}[b!]
\caption{Reordering a PQ-tree, with an interval order -- $\reorder(T,\wlt)$} \label{alg:interval_reorder}
\begin{algorithmic}[1]
\REQUIRE A PQ-tree $T$ and an interval order $\wlt$ with a sorted representation. 
\ENSURE A reordering $T'$ of $T$ such that $<_{T'}$ extends $\wlt$ if it exists.
\medskip

\STATE Calculate the handles for each individual leaf of $T$ and initiate an empty list for each endpoint.
\STATE Process the nodes of $T$ from the bottom to the root:
\FOR{a processed node $N$}
	\STATE Compute the handles of $N$ using~(\ref{eq:handle_computation}).
	\STATE Add the node $N$ to the lists of the two endpoints which are the handles of $N$.
\ENDFOR
\STATE Iterate the sorted representation and construct all orderings $\widetilde\lessdot$ for all
inner nodes $N$.
\medskip
\STATE Again process the nodes from the bottom to the root:
\FOR{a processed node $N$ with the ordering $\widetilde\lessdot$}
	\IF{the node $N$ is a P-node}
		\STATE Find any topological sort by removing minimal elements from $\widetilde\lessdot$.
		\STATE If it exists, reorder $N$ according to it, otherwise output ``no''.
	\ELSIF{the node $N$ is a Q-node}
		\STATE Test whether the current ordering or its reversal are topological sorts.
		\STATE Process the prescribed ordering from left to right, check for every element whether it is
		minimal and remove its handles from $\widetilde\lessdot$.
		\STATE If at least one ordering is correct, reorder the node, otherwise output ``no''.
	\ENDIF
\ENDFOR
\medskip
\RETURN A reordering $T'$ of $T$.
\end{algorithmic}
\end{algorithm}

Now we test for each inner node of $T$ with its ordering $\widetilde\lessdot$ whether its subtrees
can be reordered according to $\wlt$. The algorithm is correct since it works in the same way as in
Proposition~\ref{prop:general_reorder}, based on Lemma~\ref{lem:comparing_lemma}
and~\ref{lem:minimal_elements}.

Concerning the time complexity, we already discussed that we are able to compare sets of intervals
using handles in constant time, by Lemma~\ref{lem:comparing_lemma}. The precomputation of all
orderings $\widetilde\lessdot$ takes time $\O(e)$. We spend time $\O(k)$ in each node with $k$
children. Thus the total time complexity of the algorithm is linear in the size of the tree, which
is $\O(e)$.
\end{proof}

For a pseudocode, see Algorithm~\ref{alg:interval_reorder}. We note that when the orderings
$\widetilde\lessdot$ are constructed for all inner nodes, we do not need to process the tree
from the bottom to the top. We can process them independently in parallel and a reordering $T'$ of $T$
exists if and only if we succeed in reordering every inner node. 

\section{Extending Interval Graphs} \label{sec:extension_algorithm}

In this section, we describe an algorithm solving $\ext(\int)$ in time~$\O(n+m)$ which uses
Proposition~\ref{prop:interval_reorder} as a subroutine. Unlike in \S\ref{sec:pq_tree_reordering},
the interval representations consist of closed intervals. We allow the intervals to share the
endpoints and to have zero lengths. First, we describe a recognition algorithm for interval graphs
based on PQ-trees. Then we show how to modify this approach to solve $\ext(\int)$, using the
reordering algorithms from \S\ref{sec:pq_tree_reordering}.

\subsection{Recognition using PQ-trees} \label{subsec:int_recog}

Recognition of interval graphs in linear time was a long-standing open
problem, first solved by Booth and Lueker~\cite{PQ_trees} using PQ-trees. Nowadays, there are three
main approaches to linear-time recognition. The first one finds a feasible ordering of the
maximal cliques which can be done using PQ-trees. The second one uses surprising properties of the
lexicographic breadth-first search, searches through the graph several times and constructs a
representation if the graph is an interval graph~\cite{LBFS_int}. The third
one~\cite{finding_lb_graphs} tests whether the graph contains one of the minimal forbidden
subgraphs, characterized by Lekherkerker and Boland~\cite{lb_graphs}.

We describe the PQ-tree approach in details. Recall the PQ-trees from
\S\ref{sec:pq_tree_reordering}.

\heading{Maximal Cliques.} The PQ-tree approach is based on the following characterization of
interval graphs, due to Fulkerson and Gross~\cite{maximal_cliques}. A linear ordering of the maximal
cliques is called a \emph{consecutive ordering of the maximal cliques} if for every vertex the
cliques containing this vertex appear consecutively in this ordering.

\begin{lemma}[Fulkerson and Gross] \label{lem:int_char}
A graph is an interval graph if and only if there exists a consecutive ordering of the maximal
cliques.
\end{lemma}

\begin{figure}[b!]
\centering
\includegraphics{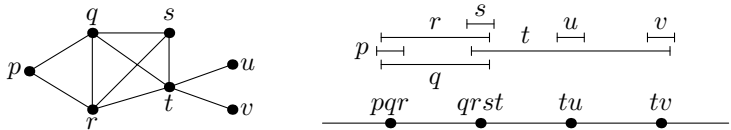}
\caption{An interval graph and one of its representations with denoted clique-points.}
\label{fig:clique_ordering}
\end{figure}

Consider an interval representation of an interval graph. For each maximal clique, consider the
intervals representing the vertices of this maximal clique and select a point in their intersection.
(We know that this intersection is non-empty because intervals of the real line have the Helly
property.) We~call these points \emph{clique-points}. For an illustration, see
Figure~\ref{fig:clique_ordering}.  The ordering of the clique-points from left to right gives the
ordering required by Lemma~\ref{lem:int_char}.  Every vertex appears in consecutive maximal cliques
since it is represented by an interval. For a maximal clique $a$, we denote the assigned
clique-point by $\cp(a)$. 

On the other hand, given a consecutive ordering of the maximal cliques, we place clique-points in this ordering
on the real line. Each vertex is represented by the interval containing exactly the clique-points of
the maximal cliques containing this vertex. Since the ordering of maximal clique is consecutive,
we obtain a valid interval representation of the graph.

The following simple lemma is useful later in proving Proposition~\ref{prop:oncas}:

\begin{lemma} \label{lem:consecutive_subgraph}
Let $<$ be a consecutive ordering of the maximal cliques and $S$ be a connected induced subgraph.
Then the maximal cliques containing at least one vertex of $S$ appear consecutively in $<$.
\end{lemma}

\begin{proof}
Consider the interval representation given by $<$ with some choice of clique-points. The union $U$
of the intervals of $S$ is a closed intervals, so it is connected. For every clique $a$, its
clique-point $\cp(a)$ is placed on $U$ if and only if $a$ contains at least one vertex from $S$.
Therefore the set of maximal cliques containing at least one vertex of $S$ appears consecutively,
with the remaining cliques on one side or the other. 
\end{proof}

\heading{Recognition Algorithm.} Every chordal graph has at most $\O(n)$ maximal cliques of total
size $\O(n+m)$ and they can be found in linear time~\cite{recog_chordal_graphs}. Since every interval
graph is chordal, we run this subroutine. If it fails, the input graph is not an interval graph, and
the recognition algorithm outputs ``no''.

According to Lemma~\ref{lem:int_char}, we want to test whether a consecutive ordering of the maximal
cliques exists. We can reduce this to the consecutive ordering problem from
\S\ref{sec:pq_tree_reordering}.  The elements $E$ are the maximal cliques of the graph.  For each
vertex $v$, we introduce the restricting set $S_v$ containing all the maximal cliques containing
this vertex $v$. Using PQ-trees, we can find a consecutive ordering of the maximal cliques and
recognize an interval graph in time $\O(n+m)$.

\subsection{Modification for \ext}

We first sketch the algorithm. We construct a PQ-tree $T$ for the input graph independently of the
partial representation. The partial representation gives another restriction---an interval order
$\wlt$ of the maximal cliques.  Using Proposition~\ref{prop:interval_reorder}, we try to find a
reordering $T'$ of the PQ-tree $T$ compatible with $\wlt$ in time $\O(n+m)$. We are going to prove
the following statement: \emph{The partial representation is extendible if and only if the
reordering subroutine succeeds.}

Since our proof is constructive, we can use it to build a representation $\calR$ extending the partial
representation $\calR'$. We place clique-points on the real line according to the ordering
$<_{T'}$. We~need to be more careful in this step. Since several intervals are pre-drawn, we cannot
change their representations, so the clique-points have to be placed correctly. Using the
clique-points, we construct the remaining intervals in a similar manner as in
Figure~\ref{fig:clique_ordering}.

Now, we describe everything in detail.

\heading{Restricting Clique-points.} Suppose that there exists a representation $\calR$ extending
$\calR'$. Then $\calR$ gives some ordering $<$ of clique-points from left to right. We want to
show that pre-drawn intervals partially specify the positions of some clique-points and give some
necessary condition for $<$.

\begin{figure}[t!]
\centering
\includegraphics{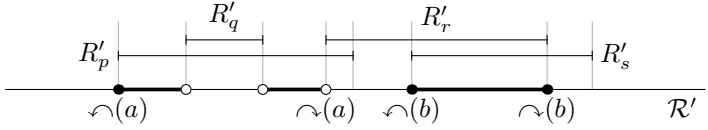}
\caption{The partial representation $\calR'$ consisting of four pre-drawn intervals. Clique-points
$\cp(a)$ and $\cp(b)$, having $P(a) = \{p\}$ and $P(b) = \{r,s\}$, can be placed to the bold parts
of the real lines.}
\label{fig:parts_of_line}
\end{figure}

For a maximal clique $a$, let $P(a)$ denote the set of all pre-drawn intervals that are
contained in $a$. Then $P(a)$ restricts the possible position of $\cp(a)$ on only those points $x$ of the
real line which are covered in $\calR'$ by exactly the pre-drawn intervals of $P(a)$ and no others.
We denote by $\clql(a)$ (resp.~$\clqr(a)$) the leftmost (resp.~the rightmost) point where the
clique-point $\cp(a)$ can be placed, formally:
\begin{eqnarray*}
\clql(a) &=& \inf\hphantom{\sup}\!\!\!\!\!\!\!
		\bigl\{ x \mid \hbox{the clique-point $\cp(a)$ can be placed on $x$}\bigr\},\\
\clqr(a) &=& \sup\hphantom{\inf}\!\!\!\!\!\!\!
		\bigl\{ x \mid \hbox{the clique-point $\cp(a)$ can be placed on $x$}\bigr\}.
\end{eqnarray*}
Notice that $(\clql(a),\clqr(a))$ is a subinterval of $\bigcap_{u \in P(a)} R'_u$. For an example,
see Figure~\ref{fig:parts_of_line}.

For a clique-point $\cp(a)$, the structure of all $x$ where $\cp(a)$ can be placed is simple.  The
endpoints of the pre-drawn intervals split the real line into several open intervals and each such
interval is called a \emph{part}. For example in Figure~\ref{fig:parts_of_line}, we have from left
to right 9 parts separated by gray lines. A clique-point $\cp(a)$ can be placed only on those parts
which contain exactly the intervals of $P(a)$ and no other pre-drawn intervals. So the set of all
points $x$ where $\cp(a)$ can be placed is a union of parts.

Also notice that the definition of $\clql(a)$ and $\clqr(a)$ does not imply that $\cp(a)$ can be
placed to all the points between $\clql(a)$ and $\clqr(a)$. If a clique-point cannot be placed
at~all, the given partial representation is clearly not extendible.

\heading{The Interval Order $\wlt$.} For two maximal cliques $a$ and $b$, we define $a \wlt b$ if
$\clqr(a) \le \clql(b)$. The definition of $\wlt$ is quite natural since $a \wlt b$ implies that
every extending representation $\calR$ has to place $\cp(a)$ to the left of $\cp(b)$. For example,
the maximal cliques $a$ and $b$ in Figure~\ref{fig:parts_of_line} satisfy $a \wlt b$.

The goal of this section is to characterize extendible partial representations as those
representations having a consecutive orderings of maximal cliques which extends $\wlt$. Consequently
for the algorithm for $\ext(\int)$, we just solve the reordering problem of the PQ-tree for $\wlt$ from
\S\ref{sec:pq_tree_reordering}. To get the linear-time complexity, we cannot use the reordering
algorithm for general partial orderings; the example in Figure~\ref{fig:quadratic_example} can be
easily generalized to get quadratically many comparable pairs in $\wlt$. Luckily, we can apply the
second reordering algorithm for interval orders:

\begin{lemma} \label{lem:clique_give_interval_order}
The relation $\wlt$ is an interval order.
\end{lemma}

\begin{proof}
The intervals representing $\wlt$ correspond to the maximal cliques of $G$. To a maximal clique $a$,
we assign an open interval $I_a = (\clql(a), \clqr(a))$. The definition of $\wlt$ exactly states
that $a \wlt b$ if and only if the intervals $I_a$ and $I_b$ are disjoint and $I_a$ is on the left
of $I_b$.
\end{proof}

The reader might be wondering why we represent interval graphs by closed intervals, but interval
orders by open intervals. The standard definition of interval graph uses closed intervals. In every
interval representation, its clique-points are strictly ordered from left to right, with no two
sharing their positions. So even when $\clqr(a) = \clql(b)$, the clique-point $\cp(a)$ is always placed
to the left of $\cp(b)$. Therefore it is natural to represent the interval orders $\wlt$ by open
intervals.

\begin{lemma} \label{lem:computing_wlt}
For a sorted partial representation $\calR'$, we can compute the sorted representation of $\wlt$ in
time $\O(n+m)$.
\end{lemma}

\begin{proof}
We sweep the real line from left to right and compute the sorted representation of $\wlt$.  As
stated above, the interval graph has $\O(n)$ maximal cliques containing in total $\O(n+m)$ vertices.
We compute for every pre-drawn vertex the list of the maximal cliques containing it, and for every
maximal clique $a$ the number $|P(a)|$ of pre-drawn vertices it contains. We initiate an empty list
$W$, a counter $i$ of pre-drawn intervals covering the currently sweeped point.

When sweeping, there are two types of events. If we encounter a set of endpoints of pre-drawn
intervals sharing a point, we first process the left endpoints, then we update $\clql$ and
$\clqr$ for this point,\footnote{We also need to update here since it might happen that the interval
$(\clql(a),\clqr(a))$ is empty for some maximal clique $a$. This can happen only if some
pre-drawn interval of $P(a)$ is a singleton.} and then we process the right endpoints. If we
sweep over a part, we just update $\clql$ and $\clqr$. In details, we do the following:
\begin{packed_itemize}
\item \emph{If we encounter a left endpoint} $\ell_u$, then we increase the counter $i$. For every
clique $a$ containing $u$, we increase its counter. If some clique $a$ has all pre-drawn intervals
placed over $\ell_u$, we add $a$ into the list $W$ of watched cliques.
\item \emph{If we encounter a right endpoint} $r_u$, we decrease the counter of pre-drawn
intervals. We ignore all maximal cliques containing $u$ till the end of the procedure, and
naturally we also remove them from $W$ if there are any.
\item \emph{The update of $\clql$ and $\clqr$} is done for all cliques $a \in W$ such that $|P(a)| = i$.
Notice that we currently sweep over exactly $i$ pre-drawn intervals, and therefore we have to sweep
over exactly the pre-drawn intervals of $P(a)$. We update $\clql(a)$ to the current point or the
infimum of the current part if it is not yet initialized. And we update $\clqr(a)$ to the supremum.
\end{packed_itemize}
In the end, we output the computed $\clql$ and $\clqr$ naturally sorted from left to right. If for
some maximal clique $a$, the value $\clql(a)$ was not initialized, the clique-point $\cp(a)$ cannot be
placed and the procedure outputs ``no''.

The procedure is clearly correct, and it remains to argue that it can be implemented in linear time.
We have the cliques in $W$ partitioned according to $|P(a)|$. When we sweep over $i$ pre-drawn
intervals, there is no $a \in W$ such that $|P(a)| > i$, and if $a,b \in W$ such that
$|P(a)|=|P(b)|=i$, then necessarily $P(a) = P(b)$. But then $\clql(a) = \clql(b)$ and $\clqr(a) =
\clqr(b)$, so we can ignore $b$ for the rest of the sweep procedure and set the values $\clql(b)$
and $\clqr(b)$ according to the clique $a$ in the end. Thus each update costs $\O(1)$. This
implementation clearly runs in $\O(n+m)$.
\end{proof}

The following proposition is the main structural result of this paper and it generalizes the
characterization of Fulkerson and Gross~\cite{maximal_cliques} to partially represented interval
graphs:

\begin{proposition} \label{prop:oncas}
A partial representation $\calR'$ is extendible if and only if there exists a consecutive ordering
of the maximal cliques extending the interval order $\wlt$.
\end{proposition}

\begin{proof}
A representation $\calR$ extending $\calR'$ gives some consecutive ordering of the maximal cliques.
It is easy to observe that the constraints given by $\wlt$ are necessary, so this consecutive
ordering has to extend $\wlt$. It remains to show the other implication.

Suppose that we have a consecutive ordering $<$ of the maximal cliques which extends $\wlt$.
We construct a representation $\calR$ extending $\calR'$ as follows. We place the
clique-points according to $<$ from left to right, always greedily as far to the left as possible.
Suppose we want to place a clique-point $\cp(a)$. Let $\cp(b)$ be the last placed clique-point.
Consider the infimum over all the points where the clique-point $\cp(a)$ can be placed and that are
to the right of the clique-point $\cp(b)$. If there is a single such point on the right of $\cp(b)$
(equal to the infimum), we place $\cp(a)$ there. Otherwise $\clql(a) < \clqr(a)$ and we place the
clique-point $\cp(a)$ to the right of this infimum by an appropriate epsilon, for example the length
of the shortest part (see the definition of $\wlt$) divided by $n$.

We prove by contradiction that this greedy procedure cannot fail; see Figure~\ref{greedy_placing}.
Let $\cp(a)$ be the clique-point for which the procedure fails. Since $\cp(a)$ cannot be placed,
there are some clique-points placed on the right of $\clqr(a)$ (or possibly on $\clqr(a)$ directly).
Let $\cp(b)$ be the leftmost one of them. If $\clql(b) \ge \clqr(a)$, we obtain $a \wlt b$ which
contradicts $b < a$ since $\cp(b)$ was placed before $\cp(a)$. So, we know that $\clql(b) <
\clqr(a)$. To get contradiction, we question why the clique-point $\cp(b)$ was not placed on the
left of $\clqr(a)$.

\begin{figure}[t!]
\centering
\includegraphics{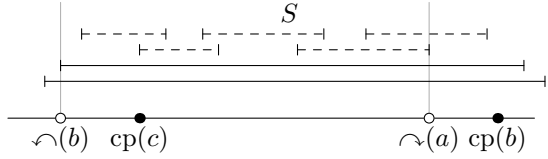}
\caption{An illustration of the proof: The positions of the clique-points $\cp(b)$ and $\cp(c)$, the
intervals of $S$ are dashed.}
\label{greedy_placing}
\end{figure}

The clique-point $\cp(b)$ was not placed on the left of $\clqr(a)$ because all these positions were either
blocked by some other previously placed clique-points, or they are covered by some pre-drawn
interval not in $P(b)$. There is at least one clique-point placed to the right of $\clql(b)$
(otherwise we could place $\cp(b)$ to $\clql(b)$ or right next to it). Let $\cp(c)$ be the
right-most clique-point placed between $\clql(b)$ and $\cp(b)$. Every point between $\cp(c)$ and
$\clqr(a)$ has to be covered by a pre-drawn interval not in $P(b)$.  Consider the set $S$ of all the
pre-drawn intervals not contained in $P(b)$ intersecting $[c,\clqr(a)]$; depicted dashed in
Figure~\ref{greedy_placing}.

Let $C$ be the set of all maximal cliques containing at least one vertex from $S$. Since $S$ induces a
connected subgraph, according to Lemma~\ref{lem:consecutive_subgraph} all maximal cliques of $C$
appear consecutively in $<$. Now, $a$ and $c$ both belong to $C$, but $b$ does not. Since $c < b$,
then $a < b$ which contradicts our original assumption $b < a$.
\end{proof}

This characterization has the following algorithmic reformulation:

\begin{corollary} \label{cor:oncas}
A partial representation $\calR'$ is extendible if and only if a PQ-tree $T$ represents all
consecutive orderings of the maximal cliques and the problem $\reorder(T,\wlt)$ can be solved.
\end{corollary}

\heading{The Algorithm.} We solve the problem $\ext(\int)$ in the following five steps. Only the
first three steps are necessary to answer the decision problem without constructing a
representation.
\begin{packed_enum}
\item Independently of the partial representation, find all maximal cliques and construct a PQ-tree
$T$ representing all consecutive orderings of the maximal cliques.
\item Construct the sorted representation of the interval order $\wlt$ by Lemma~\ref{lem:computing_wlt}.
\item Using Proposition~\ref{prop:interval_reorder}, test whether there is a reordering $T'$ of the
PQ-tree $T$ according to $\wlt$.
\item Place the clique-points from left to right according to $<_{T'}$ on the real line, greedily as
far to the left as possible.
\item Using these clique-points, construct a representation $\calR$ extending $\calR'$.
\end{packed_enum}

\begin{algorithm}[t!]
\caption{Extending Interval Graphs -- $\ext(\int)$} \label{alg:ext_int}
\begin{algorithmic}[1]
\REQUIRE An interval graph $G$ and a partial representation $\cal R'$. 
\ENSURE A representation $\cal R$ extending $\cal R'$ if it exists.
\medskip

\STATE Compute maximal cliques and construct a PQ-tree.
\STATE Sweep $\calR'$ from left to right and construct the sorted representation of $\wlt$.
\STATE Use Algorithm~\ref{alg:interval_reorder} to reorder the PQ-tree according $\wlt$.
\STATE If any of these steps fails, no representation exists and output ``no''.
\medskip
\STATE Place the clique-points according to the ordering $<_{T'}$ from left to right:
\FOR{a clique-point $\cp(a)$ placed after $\cp(b)$}
	\STATE Compute the infimum of all points of the real line on the right of $\cp(b)$ where
			$\cp(a)$ can be placed.
	\STATE If there is single such point, place $\cp(a)$ there.
	\STATE Otherwise place $\cp(a)$ by $\eps$ on the right of the infimum, where $\eps$ is the size
			of the smallest part divided by $n$.
\ENDFOR
\STATE Construct $\cal R$ for the remaining intervals on top of the placed clique-points.
\medskip
\RETURN A representation $\cal R$ extending $\cal R'$.
\end{algorithmic}
\end{algorithm}

Step~1 is the original recognition algorithm. In Step~2, we compute splitting of the real line into
parts and construct a sorted representation of $\wlt$. In Step~3, we apply the algorithm
of Proposition~\ref{prop:interval_reorder}. Step~4 is the greedy procedure from the proof of
Proposition~\ref{prop:oncas}. In Step~5, we construct intervals representing the vertices of $G
\setminus G'$ as in Figure~\ref{fig:clique_ordering}; we construct each such interval on top of the
corresponding clique-points. See Algorithm~\ref{alg:ext_int} for a pseudocode.

Now we are ready to prove the main result of this paper, Theorem~\ref{thm:ext_int} which states that
the problem $\ext(\int)$ can be solved in time $\O(n+m)$:

\begin{proof}[Theorem~\ref{thm:ext_int}]
The correctness of the algorithm is implied by Corollary~\ref{cor:oncas}. Concerning the
complexity, the total size of all maximal cliques is at most $\O(n+m)$ and the PQ-tree can be
constructed in this time. Using Lemma~\ref{lem:computing_wlt}, we can construct the sorted
representation of $\wlt$ in time $\O(n+m)$. According to Proposition~\ref{prop:interval_reorder},
the PQ-tree can be reordered according to $\wlt$ in time $\O(n+m)$. Finally, a representation
$\calR$ extending $\calR'$ can be constructed, if necessary, in time $\O(n+m)$.
\end{proof}

\section{Simultaneous Representations of Interval Graphs} \label{sec:simultaneous_representations}

The input of the simultaneous representations problem gives several graphs $G_1,\dots,G_k$ having a
common intersection $I$. The task is to construct their representations $\calR^1,\dots,\calR^k$ which
represent the vertices of $I$ the same; see Figure~\ref{fig:simultaneous_representations}. Formally
it is the following decision problem:

\begin{figure}[b!]
\centering
\includegraphics{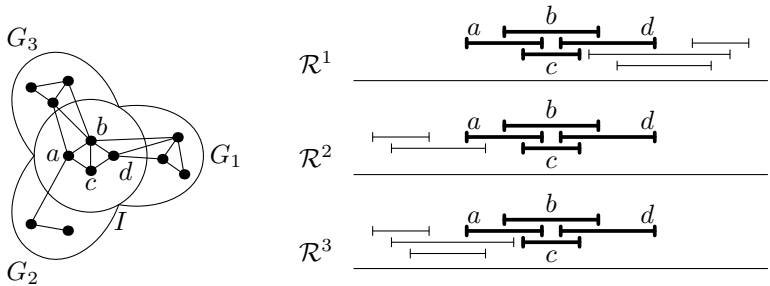}
\caption{An example of three interval graphs with simultaneous representations assigning the same
intervals to $I = \{a,b,c,d\}$.}
\label{fig:simultaneous_representations}
\end{figure}

\computationproblem
{Simultaneous representations -- $\simrep(\calC)$}
{Graphs $G_1,\dots,G_k$ such that $G_i \cap G_j = I$ for all $i \ne j$.}
{Do there exist representations $\calR^1,\dots,\calR^k$ such that
$\calR^i = \bigl\{R^i_u \mid u \in V(G_i)\bigr\}$ represents $G_i$ and for every
$u \in I$ we have $R^i_u = R^j_u$ for every $i$ and $j$.}

Jampani et al.~\cite{jampani2} show that for permutation and comparability graphs the corresponding
problems can be solved in polynomial time for any number of graphs, and for chordal graphs the
problem is polynomially solvable for $k=2$ and \cNP-complete when $k$ is a part of the input. For
two interval graphs, the paper~\cite{jampani} gives an $\O(n^2 \log n)$ algorithm which Bl\"asius
and Rutter~\cite{blas_rutter} improve to $\O(n+m)$. For circle graphs, the problem is \cNP-complete
when $k$ is a part of the input~\cite{cfk} and open even for $k=2$.

\heading{Relation to \ext.} For many classes, the simultaneous representations problem is closely
related to the partial representation extension problem.  A negative example is the class of chordal
graphs, denoted by \chor. The problem $\simrep(\chor)$ is polynomially solvable for
$k=2$~\cite{jampani2}, but $\ext(\chor)$ is \cNP-complete~\cite{kkos}.  On the other hand, we get
the following relations for interval graphs.

We sketch an easy reduction of $\ext(\int)$ to $\simrep(\int)$ for $k=2$ of Bl\"asius and Rutter,
see~\cite[Section 4.1]{blas_rutter}. As the graph $G_1$, we put the input graph $G$, and as $I$ we
put the pre-drawn vertices $V(G')$. Now consider $\calR'$, add a path going from left to right
consisting of short intervals. This represents some interval graph which we put as $G_2$. The key is
that $G_2$ fixes any representation of $I$ to be topologically equivalent to $\calR'$. So the
question whether $G_1$ can be simultaneously represented with $G_2$ is equivalent to $\ext(\int)$.
The linear-time algorithm for $\ext(\int)$ of Bl\"asius and Rutter~\cite{blas_rutter} is based on
this reduction.

The other relation is that if $I$ is small enough, we can use the partial representation extension
algorithm to test all possible representations of $I$.  As a straightforward corollary of
Theorem~\ref{thm:ext_int}, the problem $\simrep(\int)$ is \cFPT\ in the size of $I$: 

\begin{corollary}
The problem $\simrep(\int)$ can be solved in $\O((n+m) (2\ell)!)$ where $\ell =
|I|$, $n = |V(G_1)|+\cdots+|V(G_k)|$, and $m = |E(G_1)|+\cdots+|E(G_k)|$.
\end{corollary}

\begin{proof}
There are $(2\ell)!$ different representations of $I$, given by all possible orderings of the $2\ell$
endpoints. (Indeed, many of these orderings do not give a correct representation of $I$.) For each
representation $\calR'$ of $I$, we test whether it is extendible to representations
$\calR^1,\dots,\calR^k$. This can be done by running $k$ instances of the algorithm of
Theorem~\ref{thm:ext_int} which takes the total time $\O(n+m)$. Since we need to test $(2\ell)!$
possible representations, the total time is $\O((n+m)(2\ell)!)$.

For the correctness, if the algorithm succeeds in constructing $\calR^1,\dots,\calR^k$, the
simultaneous representations problem is solvable. On the other hand, if the simultaneous
representations problem is solvable, there exists some common representation of $I$ and we test a
representation $\calR'$ which is topologically equivalent to it. Since the solvability of partial
representation extension depends only on the ordering of the endpoints, then the representation
$\calR'$ is extendible to $\calR^1,\dots,\calR^k$.
\end{proof}

We note that the same idea works for several other graph classes. For instance, Chaplick et
al.~\cite{cfk} give the same \cFPT\ algorithm for simultaneous representations of circle graphs
based on partial representation extension. Further, they prove that the simulateneous
representations problem is \cNP-complete when $k$ is a part of the input. In the case of interval
graphs, the complexity of simultaneous representations is open when $k$ is a part of the input.

\section{An Open Problem} \label{sec:open_problem}

We conclude the paper with currently the main open problem. \emph{Circular-arc graphs} (\ca) are
intersection graphs of arcs of a circle; see~\cite{egr}.

\begin{problem}
Can the problem $\ext(\ca)$ be solved in polynomial time?
\end{problem}

Solving this problem might lead to a better understanding of the class itself. All known
polynomial-time recognition algorithms are quite complex and construct specific types of
representations called \emph{canonical representations}; see~\cite{hsu_ca,linear_ca}. To solve
$\ext(\ca)$, the structure of all representations needs to be better understood which could lead to
a major breakthrough concerning this and other classes.

\section*{Acknowledgements}

We are very thankful to Pavol Hell for suggesting the PQ-trees approach, and to Martin Balko and
Ji\v{r}\'{i} Fiala for comments concerning writing.

\bibliographystyle{siam}
\bibliography{extending_int_journal}

\begin{thebibliography}{10}

\bibitem{angelini}
{\sc P.~Angelini, G.~D. Battista, F.~Frati, V.~Jel\'{\i}nek,
  J.~Kratochv\'{\i}l, M.~Patrignani, and I.~Rutter}, {\em Testing planarity of
  partially embedded graphs}, in Proceedings of the 21st Annual ACM-SIAM
  Symposium on Discrete Algorithms, SODA'10, 2010, pp.~1030--1043.

\bibitem{bko}
{\sc M.~Balko, P.~Klav\'{\i}k, and Y.~Otachi}, {\em Bounded representations of
  interval and proper interval graphs}, in Algorithms and Computation,
  vol.~8283 of LNCS, Springer, 2013, pp.~535--546.

\bibitem{benzer_interval_graphs}
{\sc S.~Benzer}, {\em On the topology of the genetic fine structure}, Proc.
  Nat. Acad. Sci. U.S.A., 45 (1959), pp.~1607--1620.

\bibitem{blas_rutter}
{\sc T.~Bl{\"a}sius and I.~Rutter}, {\em Simultaneous {PQ}-ordering with
  applications to constrained embedding problems}, in Proceedings of the 24th
  Annual ACM-SIAM Symposium on Discrete Algorithms, SODA'13, 2013,
  pp.~1030--1043.

\bibitem{PQ_trees}
{\sc K.~S. Booth and G.~S. Lueker}, {\em Testing for the consecutive ones
  property, interval graphs, and planarity using {PQ}-tree algorithms}, J.
  Comput. System Sci., 13 (1976), pp.~335--379.

\bibitem{cfk}
{\sc S.~Chaplick, R.~Fulek, and P.~Klav\'{\i}k}, {\em Extending partial
  representations of circle graphs}, in Graph Drawing, vol.~8242 of LNCS,
  Springer, 2013, pp.~131--142.

\bibitem{LBFS_int}
{\sc D.~G. Corneil, S.~Olariu, and L.~Stewart}, {\em The {LBFS} structure and
  recognition of interval graphs}, SIAM Journal on Discrete Mathematics, 23
  (2009), pp.~1905--1953.

\bibitem{fishburn}
{\sc P.C. Fishburn}, {\em Interval orders and interval graphs: a study of
  partially ordered sets}, Wiley, 1985.

\bibitem{maximal_cliques}
{\sc D.~R. Fulkerson and O.~A. Gross}, {\em Incidence matrices and interval
  graphs.}, Pac. J. Math., 15 (1965), pp.~835--855.

\bibitem{agt}
{\sc M.~C. Golumbic}, {\em Algorithmic Graph Theory and Perfect Graphs},
  North-Holland Publishing Co., 2004.

\bibitem{hajos}
{\sc G.~Haj{\'o}s}, {\em {\"U}ber eine {A}rt von {G}raphen}, Internationale
  Mathematische Nachrichten, 11 (1957), p.~65.

\bibitem{hsu_ca}
{\sc W.~Hsu}, {\em \$o(m.n)\$ algorithms for the recognition and isomorphism
  problems on circular-arc graphs}, SIAM J. Comput., 24 (1995), pp.~411--439.

\bibitem{jampani}
{\sc K.~R. Jampani and A.~Lubiw}, {\em Simultaneous interval graphs}, in
  Algorithms and Computation, vol.~6506 of Lecture Notes in Computer Science,
  2010, pp.~206--217.

\bibitem{jampani2}
\leavevmode\vrule height 2pt depth -1.6pt width 23pt, {\em The simultaneous
  representation problem for chordal, comparability and permutation graphs},
  Journal of Graph Algortihms and Applications, 16 (2012), pp.~283--315.

\bibitem{kkkw}
{\sc P.~Klav{\'i}k, J.~Kratochv{\'i}l, T.~Krawczyk, and B.~Walczak}, {\em
  Extending partial representations of function graphs and permutation graphs},
  in Algorithms -- ESA 2012, vol.~7501 of Lecture Notes in Computer Science,
  2012, pp.~671--682.

\bibitem{kkorssv}
{\sc P.~Klav\'{\i}k, J.~Kratochv\'{\i}l, Y.~Otachi, I.~Rutter, T.~Saitoh,
  M.~Saumell, and T.~Vysko\v{c}il}, {\em Extending partial representations of
  proper and unit interval graphs}, Accepted to SWAT 2014,  (2014).

\bibitem{kkos}
{\sc P.~Klav\'{\i}k, J.~Kratochv\'{\i}l, Y.~Otachi, and T.~Saitoh}, {\em
  Extending partial representations of subclasses of chordal graphs}, in
  Algorithms and Computation -- ISAAC, vol.~7676 of Lecture Notes in Computer
  Science, 2012, pp.~444--454.

\bibitem{kkv}
{\sc P.~Klav\'{\i}k, J.~Kratochv\'{\i}l, and T.~Vysko\v{c}il}, {\em Extending
  partial representations of interval graphs}, in Theory and Applications of
  Models of Computation - 8th Annual Conference, TAMC 2011, vol.~6648 of
  Lecture Notes in Computer Science, 2011, pp.~276--285.

\bibitem{ks}
{\sc P.~Klav\'{\i}k and M.~Saumell}, {\em Minimal obstructions for partial
  representation extension of interval graphs}, In preparation,  (2014).

\bibitem{int_lengths_intersections}
{\sc J.~Kobler, S.~Kuhnert, and O.~Watanabe}, {\em Interval graph
  representation with given interval and intersection lengths}, in Algorithms
  and Computation, vol.~7676 of Lecture Notes in Computer Science, 2012,
  pp.~517--526.

\bibitem{lb_graphs}
{\sc C.~Lekkerkerker and D.~Boland}, {\em Representation of ﬁnite graphs by a
  set of intervals on the real line}, Fund. Math., 51 (1962), pp.~45--64.

\bibitem{finding_lb_graphs}
{\sc N.~Lindzey and R.~M. McConnell}, {\em On finding tucker submatrices and
  lekkerkerker-boland subgraphs}, in Graph-Theoretic Concepts in Computer
  Science, vol.~8165 of Lecture Notes in Computer Science, 2013, pp.~345--357.

\bibitem{every_graph_is_an_intersection_graph}
{\sc E.~S. Marczewski}, {\em Sur deux propri\'et\'es des classes d'ensembles},
  Fund. Math., 33 (1945), pp.~303--307.

\bibitem{linear_ca}
{\sc R.~M. McConnell}, {\em Linear-time recognition of circular-arc graphs},
  Algorithmica, 37 (2003), pp.~93--147.

\bibitem{tig}
{\sc T.~A. McKee and F.~R. McMorris}, {\em Topics in Intersection Graph
  Theory}, SIAM Monographs on Discrete Mathematics and Applications, 1999.

\bibitem{patrignani}
{\sc M.~Patrignani}, {\em On extending a partial straight-line drawing}, in
  Lecture Notes in Computer Science, vol.~3843, 2006, pp.~380--385.

\bibitem{int_prescribed_lengths}
{\sc I.~Pe'er and R.~Shamir}, {\em Realizing interval graphs with size and
  distance constraints}, SIAM J. Discret. Math., 10 (1997), pp.~662--687.

\bibitem{roberts_theorem}
{\sc F.~S. Roberts}, {\em Indifference graphs}, Proof techniques in graph
  theory,  (1969), pp.~139--146.

\bibitem{roberts_discrete_models}
\leavevmode\vrule height 2pt depth -1.6pt width 23pt, {\em Discrete
  Mathematical Models, with Applications to Social, Biological, and
  Environmental Problems}, Prentice-Hall, Englewood Cliffs, 1976.

\bibitem{recog_chordal_graphs}
{\sc D.~J. Rose, R.~E. Tarjan, and G.~S. Lueker}, {\em Algorithmic aspects of
  vertex elimination on graphs}, SIAM Journal on Computing, 5 (1976),
  pp.~266--283.

\bibitem{egr}
{\sc J.~P. Spinrad}, {\em Efficient Graph Representations}, Field Institute
  Monographs, 2003.

\bibitem{stoffers}
{\sc K.~E. Stoffers}, {\em Scheduling of traffic lights--a new approach},
  Transportation Research, 2 (1968), pp.~199--234.

\bibitem{trotter}
{\sc W.~T. Trotter}, {\em New perspectives on interval orders and interval
  graphs}, in in Surveys in Combinatorics, Cambridge Univ. Press, 1997,
  pp.~237--286.

\end{thebibliography}
\end{document}